\documentclass[a4paper,aps,prl,twocolumn,groupedaddress,showpacs]{revtex4}
\usepackage{graphicx}
\usepackage{dcolumn}
\usepackage{bm}

\begin{document}
\pacs{73.63.Kv, 72.25.-b}
\title{Spin state mixing in InAs double quantum dots}
\author{A.~Pfund, I.~Shorubalko, K.~Ensslin and R.~Leturcq}

\affiliation{Solid State Physics Laboratory, ETH Z\"urich, 8093 Z\"urich, Switzerland\\
E-mail: leturcq@phys.ethz.ch }

\begin{abstract}
We quantify the contributions of hyperfine and spin-orbit mediated singlet-triplet mixing in weakly coupled InAs quantum dots by electron transport spectroscopy in the Pauli spin blockade regime. In contrast to double dots in GaAs, the spin-orbit coupling is found to be more than two orders of magnitudes larger than the hyperfine mixing energy. It is already effective at magnetic fields of a few mT, where deviations from hyperfine mixing are observed.
\end{abstract}

\maketitle
Spin dependent interactions such as spin-orbit (SO) interaction and hyperfine (HF) coupling to the nuclei have significant influence on spin transport in solid state devices. The perspective of active control of these mechanisms stimulated many proposals for ``spintronic'' devices \cite{wolf2001,datta_das_SFET}.
Spin states in coupled semiconductor quantum dots are considered as possible realizations of quantum bits in scalable solid state quantum computers \cite{PhysRevA.57.120}. Electrical control of SO interactions \cite{golovach:165319,PhysRevLett.97.240501-flindt} as well as dynamic coupling of electrons and nuclei \cite{KaneSiQC,ono:256803} could provide a convenient way for qubit operations. However, both effects are at the same time a major source of perturbation, since spin state mixing enables various paths of spin relaxation \cite{PhysRevB.64.195306,golovach:016601,sasaki:056803,meunier:126601}
In GaAs double quantum dots (DQDs), HF interactions have been identified to dominate the spin mixing at small magnetic fields, while SO interactions are not relevant in this regime \cite{Koppens:2005qy,johnsonPulse,Jouravlev:2006kx,PhysRevLett.88.186802}. In single quantum dots, SO is the main source for spin relaxation, especially at high magnetic fields \cite{meunier:126601,AmashaCondMat2006,golovach:016601,PhysRevB.64.125316}.

These properties are specific for the considered material. Spin-orbit interactions and the coupling to magnetic fields are expected to be orders of magnitudes stronger in InAs compared to GaAs. The interplay of relaxation processes mediated by SO and HF interaction can lead to an effectively suppressed spin relaxation, which requires strongly coupled dots \cite{pfund-2007-DNSP}. In order to quantify the relevant energy scales, we focus on the weakly coupled regime. Here, transport occurs for aligned levels and the mixing energies can be analyzed in detail.

We investigate mixing of singlet (S) and triplet (T) states in an InAs double quantum dot. Electron transport spectroscopy in the Pauli spin-blockade regime of the weakly coupled dots allows to identify the relevant spin states. We are able to distinguish the contributions of SO and HF coupling for small and large magnetic fields. Similar to recent experiments in single InAs dots \cite{fuhrer-SO-2007}, we observe a strong SO-induced S-T mixing at large magnetic fields corresponding to a coupling energy of $\Delta_{SO}=0.2\,$meV. The mixing energy by the uncorrelated hyperfine fields in the two dots is found to be close to three orders of magnitude smaller for small external fields. In contrast to GaAs DQDs, we observe clear deviations from the previously studied \cite{Koppens:2005qy,Jouravlev:2006kx} HF mixing already at millitesla fields, which can be attributed to SO interactions. The results allow to quantify the experimental parameters required for electrical field induced spin manipulation \cite{golovach:165319,PhysRevLett.97.240501-flindt}.

\begin{figure}
\includegraphics[width=0.45\textwidth]{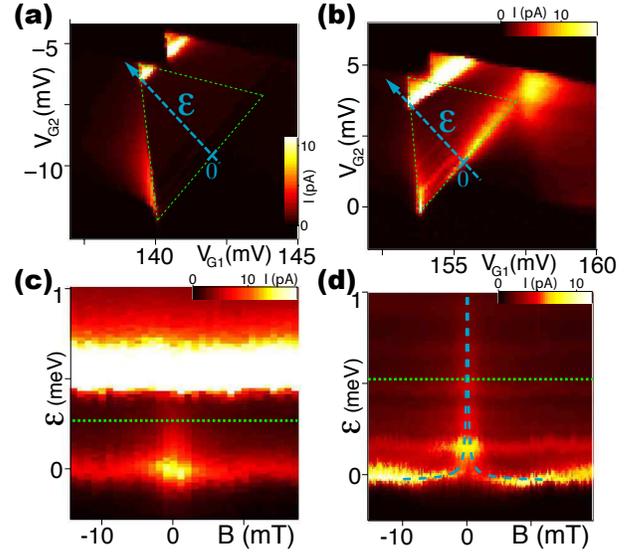}
\caption{a) Double-dot current $I_{SD}$ for $V_{SD}=2\,$mV as a function of gate voltages $V_{G1}$,$V_{G2}$. The coupling gate voltage is fixed at $V_{GC}=-120\,$mV. Spin blockade suppresses current in the base region of the triangle. Along the dashed line, the dot levels are detuned by an energy $\varepsilon$. b) Same for weaker coupling $V_{GC}=-170\,$mV ($V_{SD}=1.6\,$mV). c) Plot of $I_{SD}$ in dependence of detuning $\varepsilon$ and magnetic field $B$ for $V_{GC}=-180\,$mV. d) Same for $V_{GC}=-170\,$mV. Dashed lines are fits to a model for tunnel-coupled levels $(t=6\,\mu$eV, $g^*=7$), see text.}
\end{figure}

The device is fabricated in an InAs nanowire, catalytically grown by metal organic vapor phase epitaxy. The wire diameter is around $100\,$nm. Three top-gates of $70\,$nm width and separation define a serial DQD along the nanowire (NW) \cite{pfund:252106} (see inset of Fig.\,2). We refer to the two outer gates as $G1,G2$ (tuning energy levels in dot 1 and 2) and to the center gate as $GC$ (tuning the inter-dot coupling). Transport measurements were performed in a dilution refrigerator at an electronic temperature of $\sim100\,$mK and with a magnetic field aligned perpendicular to the nanowire axis.

We could tune the three gates to create two weakly coupled quantum dots. The states can then be labeled by the occupation numbers (n,m) for dot 1 and 2. For finite bias voltage $V_{SD}$, electron transport is forbidden due to Coulomb blockade everywhere except for triangular regions in the $V_{G1}$-$V_{G2}$-plane  \cite{vanderWiel01} (dashed lines in Figs.\,1(a,b)). Here, the dot states are in the bias window and sequential transport through the serial dots is possible.
In Fig.\,1(a), a measurement of the current $I_{SD}$ through the DQD is shown for $V_{SD}=2\,$mV. Transport is strongly suppressed in the base region of the triangle, but not at the corner points and side edges. This can be explained by Pauli spin-blockade (SB) \cite{Ono_rectification}. Considering the DQD in an initial $(0,1)$-state, a second electron can be loaded into either the singlet $S(1,1)$ or a $(1,1)$-triplet (named $T_m(1,1)$ with $m=0,\pm1$ according to the $z$-component of the spin). The ground state of the $(0,2)$-configuration at zero B-field is a singlet. Sequential transport is therefore blocked due to spin conservation, once the second electron entered the DQD in a $(1,1)$-triplet. This SB is lifted, if the mutual detuning $\varepsilon$ of the states in the two dots exceeds the $(0,2)$ singlet-triplet splitting $J_{02}$, which gives rise to the strong current in the tip of the triangle (Fig.\,1(a)). We could not determine the absolute number of electrons in the DQD in this experiment. However, SB was also observed in GaAs DQD containing many electrons \cite{johnson:165308}, if the core electrons form inert spin pairs. In InAs this is likely to occur, because exchange interactions are expected to be weak due to the small effective mass. In agreement with this assumption, we observed SB only when changing the total number of electrons by two (not shown).

In Fig.\,1(b), the same triangle is shown for weaker inter-dot coupling. Compared to Fig.\,1(a), the splitting $J_{02}$ is decreased due to the change of the confinement \cite{0034-4885-64-6-201}. Similar to experiments in GaAs \cite{Koppens:2005qy}, a large current occurs around detuning $\varepsilon=0$. 

In the following, we study $I_{SD}$ as a function of magnetic field and level detuning $\varepsilon$. The gate changes have been transformed into energy using the leverarms obtained by relating the size of the triangles to the bias voltage \cite{vanderWiel01}.
Figures 1(c) and (d) show measurements for two different center gate settings, but both still corresponding to weak coupling. Similar to \cite{Koppens:2005qy}, the large current for finite $\varepsilon$ is reduced already at fields of a few mT. In the case of slightly stronger coupling (Fig.\,1(d)), the peak splits into characteristic wings around $B=0$ that merge at higher detuning $\varepsilon$.

The baseline corresponding to $\varepsilon=0$ at small B cannot be suppressed completely even for large fields. The continuation of Fig.\,1(d) up to $B=5$\,T is shown in Fig.\,2. The baseline shifts linearly between $0\,$T and $2\,$T. The top peak, separated by $J_{02}=1.1\,$meV at $B=0$, splits into three branches with very different current levels (dashed lines). At $B\approx 2.7\,$T, a pronounced anticrossing of the two lowest peaks occurs, similar to recent observations in InAs single quantum dots \cite{fuhrer-SO-2007}.

\begin{figure}
\includegraphics[width=0.45\textwidth]{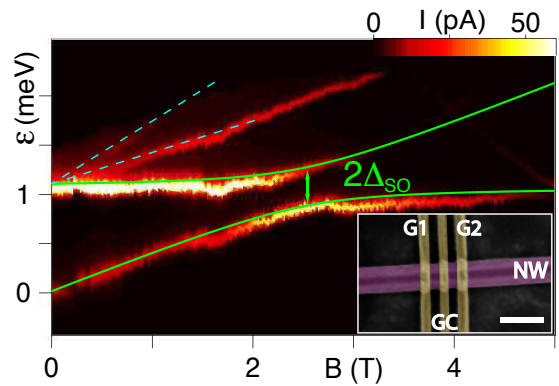}
\caption{Current as a function of detuning $\varepsilon$ and magnetic field $B$ as in Fig.\,1(d), but on a larger field scale. The lines show fits to the model described in the text $(\Delta_{SO}=200\,\mu$eV, $g^*=7$, $g_{2}^*=5.5$). An overal shift of \mbox{$0.3\,$ meV/T} has been removed, which could be due to orbital effects. Inset: Scanning electron micrograph of a representative device. Ti/Au top-gates $G1$,$G2$ and $GC$ define a double quantum dot in the InAs nanowire (NW). Scalebar $200\,$nm.}
\end{figure}

For further analysis of the low field regime, we plot the detuning dependence of the relevant DQD levels around $\varepsilon\approx 0$ in Fig.\,3(c), as obtained in a Hund-Mulliken model \cite{PhysRevB.59.2070}. A tunnel coupling $t$ hybridizes the singlets $S(1,1)$ and $S(0,2)$. This is visible as an anticrossing of $\sim 2t$ between the branches $S$. An external field $B$ splits the triplets by a Zeeman energy $g^*\mu_\textrm{B}B$, where $g^*$ is the effective g-factor and $\mu_\textrm{B}$ is the Bohr magneton.

\begin{figure}
\includegraphics[width=0.45\textwidth]{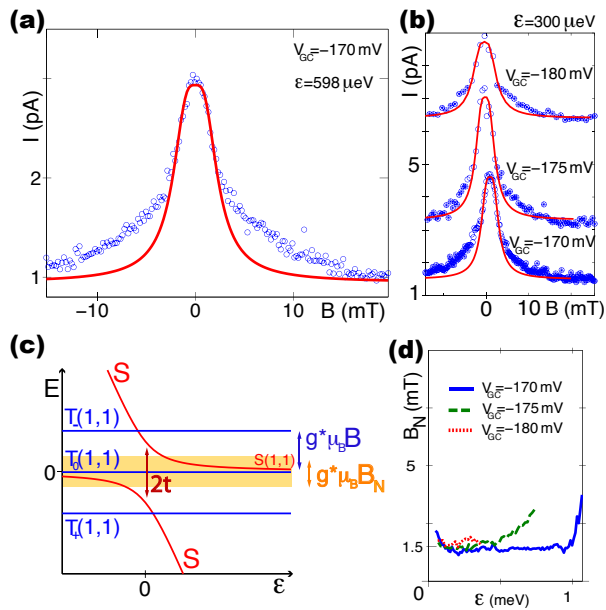}
\caption{a) Circles: Cut along the dotted green line in Fig.\,1(d). The solid red line is a fit of the center peak with Eq.\,(11) of \cite{Jouravlev:2006kx}. b) Similar traces along the dotted line in Fig.\,1(c) for 3 different values of the coupling gate $V_{GC}$. Solid red lines are fits of the center peaks. c) Level scheme of the singlets $S$ and the $(1,1)$-triplets around zero detuning $\varepsilon$. Tunnel coupling $t$ induces an anticrossing of the hybridized singlets. The Zeeman energy $g^*\mu_\textrm{B} B$ splits the triplets. A random nuclear field with amplitude $B_\textrm{N}$ mixes the states in the shaded region. d) Extracted $B_N$ from the central peak fits as a function of detuning for different couplings.}
\end{figure}

Enhanced current around $B\approx 0$ and $\varepsilon\approx 0$ could be explained with HF mixing \cite{Koppens:2005qy,Jouravlev:2006kx}. The HF coupling can be expressed by a random nuclear field as a cumulative effect of those nuclei that overlap with the electronic wavefunction \cite{PhysRevB.65.205309}. The distribution of this effective field is characterized by the width $B_N$, which is a measure for the amplitude of the field fluctuations \cite{PhysRevB.65.205309}. This mixing is efficient for the $(1,1)$ singlets and triplets where electrons are distributed over different dots.

A $(1,1)$--triplet is mixed to $S(1,1)$, if their splitting $\Delta_{ST}$ is smaller than the HF coupling, quantified by $g^*\mu_\textrm{B}B_N$. 
This regime is indicated by the shaded region around $B=0$ in Fig.\,3(c).
An external field $B$ splits the triplets $T_\pm(1,1)$ from $S(1,1)$ and SB is recovered by these states. In the limit $g^*\mu_\textrm{B}B_N > \Delta_{ST}$, this leads to the current peak around $B=0$ in Fig.\,1(c). 
For very weak coupling as in Fig.\,1(c), the anticrossing of the $S$ branches is narrow and the above condition is always fulfilled. For stronger coupling (Fig.\,1(d)), it can bee achieved at finite detuning, as obvious from Fig.\,3(c). The width of the characteristic current peak allows the determination of $B_N$ \cite{Jouravlev:2006kx}. 

In Fig.\,3(a) we plot a current trace (circles) as a a function of $B$ for fixed detuning, indicated by the dotted line in Fig.\,1(d). The complete data can not be fitted satisfactory with a single curve according to Eq.\,(11) of \cite{Jouravlev:2006kx}. This feature is consistently observed for different values of inter-dot coupling, as shown in Fig.\,3(b). However, a fit to the central peak leaving out the wide tails for $|B|>3\,$mT leads to good agreement.

From the fits of the central peaks, we extract $B_N\approx1.5\pm0.2\,$mT for the effective field amplitude. This result is almost independent on inter-dot coupling and detuning as shown in Fig.\,3(d), supporting the validity of the model in \cite{Jouravlev:2006kx} (the increase for large $\varepsilon$ can be explained by the proximity of the $(0,2)$-triplets).
The value corresponds to the HF fluctuation amplitude of $N\approx 0.5\cdot10^5$ nuclei \cite{JRGoldmanThesis}, which is consistent with the dot size evaluated from charging energy and excited state spectrum \cite{pfund:252106}.

For increased tunnel coupling $t$, the anticrossing is larger and $T_{0,\pm}(1,1)$ are separated from $S$ for small $B$ and $\varepsilon$. Increasing $B$, the triplet $T_{-}(1,1)$ is again mixed to the upper singlet branch, which has $(1,1)$-character for small detuning. This explains the splitting of the current peak around $B=0$ into wings, as observed in Fig.\,1(d). Fitting the upper singlet branch (see Fig.\,3(c)) in the vicinity of the anticrossing to the wings in Fig.\,1(d) allows to quantify the tunnel coupling $t=6\,\mu$eV (using $g^*=7$ as determined below).

In the model of \cite{Jouravlev:2006kx}, additional spin mixing mechanisms are neglected, which is justified by experimental results in GaAs DQDs \cite{Koppens:2005qy}. Since SO interaction is expected to be much stronger in InAs compared to GaAs, we suggest that the observed wide tails of the current peaks in our experiment are related to this additional contribution. Singlets and ($m=\pm1$)--triplets are also hybridized by the SO coupling. This enhances the HF-induced anticrossing of those levels. The resulting states therefore sustain a singlet contribution up to larger Zeeman splitting and hence a higher external field is required to recover spin-blockade by the $T_\pm(1,1)$ states.
We note that the field scale of the wide tails ($\sim 10\,$mT) agrees with the onset of inelastic spin relaxation for strongly coupled InAs dots in \cite{pfund-2007-DNSP}, which was related to the influence of SO interaction.

The importance of SO interaction is confirmed by the measurement for larger field and detuning (Fig.\,2). To interpret the observed current peaks and their magnetic field dependence, we extend the scheme for the detuning dependence of the DQD states as shown in Fig.\,4. At $B=0$, the $(0,2)$-triplets are separated from $S(0,2)$ by $J_{02}$. A tunnel coupling $t$ that hybridizes states with equal spin quantum numbers leads to anticrossings. For weakly coupled dots, sequential tunneling is the strongest transport path and resonant current peaks occur at those detunings $\varepsilon$, where $(1,1)$ and $(0,2)$ states are mixed by one of the described mechanisms.

The strongest current line occurs at a detuning corresponding to $J_{02}=1.1\,$meV at $B=0$. We relate this peak to tunnel mixing of $(1,1)$ and $(0,2)$ triplets with the \emph{same} spin quantum number $m=0,\pm1$. The corresponding anticrossings are labeled $X$ in Fig.\,4. If the effective g-factors $g^*$ for $T_\pm(1,1)$ and $g_2^*$ for $T_\pm(0,2)$ are close, all 3 anticrossings occur at almost the same B-independent detuning and give rise to a single peak. 

The lowest line in Fig.\,2 shifts linearly in B up to $\sim2\,$T. We explain it by probing the lower singlet branch with the state $T_+(1,1)$, which is split from $T_0(1,1)$ by $g^*\mu_\textrm{B}B$ (see Fig.\,4). At the degeneracy point of both states (labeled $Y$), HF and SO mix $S$ and $T_+$. The resonant current involves a first order spin-flip \cite{Dickmann:2003lr} and is consequently weaker than the tunnel peak $X$. From the slope, we extract an effective g-factor $g^*=7$ for the $(1,1)$-triplets.
We note that we did not compensate for a quadratic shift in $B$, which would be the expected orbital effect of the magnetic field in single dots \cite{PhysRevLett.77.3613}.

\begin{figure}
\includegraphics[width=0.45\textwidth]{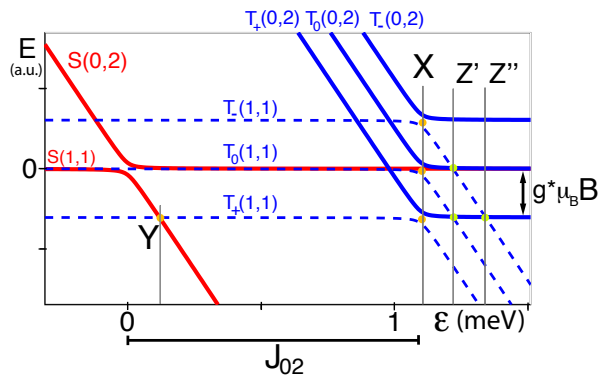}
\caption{Energies of the relevant states in the spin-blockade regime as a function of detuning $\varepsilon$. $J_{02}$ denotes the $(0,2)$ singlet-triplet splitting. A tunnel coupling energy $t$ mixes states with the same spin quantum numbers leading to anticrossings around $\varepsilon\approx 0$ and at the points marked with $X$. In the figure, a magnetic field $B$ with $g^*\mu_\textrm{B}B\gg t$ splits the triplets. Spin-flip processes can also lead to mixing at other degeneracy points such as $Y$ and $Z^\prime,Z^{\prime\prime}$.}
\end{figure}

The two upper lines in Fig.\,2 are as well much weaker than the peak due to tunnel coupling. Comparing to Fig.\,4, we suggest that these lines arise from the degeneracies named $Z^\prime$ and $Z^{\prime\prime}$. The involved mixing processes require higher order spin flips. This is consistent with the much weaker intensity of these current peaks. This model is also supported by the g-factor $g_2^*=5.5$ extracted from the slopes of the upper lines. In contrast to $(1,1)$ triplets, the relevant states $T_\pm(0,2)$ involve excited orbital states of dot 2, which could explain the difference of $g^*$ and $g_2^*$ \cite{fuhrer-SO-2007}. To further test the consistency, we opened $G1$ and studied the B-field dependence of the excited states in the single dot 2 with the same electronic occupation. This yielded again $g_2^*=5.5$ (not shown).

If $g^*\mu_\textrm{B}B$ approaches $J_{02}$, the singlet $S(0,2)$ becomes degenerate with $T_+(0,2)$. We observe a pronounced anticrossing of the lower two lines with $2\Delta_{SO}\approx0.4\,$meV. Similar to recent measurements in InAs single dots \cite{fuhrer-SO-2007}, this can be explained by SO interactions. Contributions of the HF interaction are expected to be negligible in this situation, since electrons in $(0,2)$ states experience the same nuclear field \cite{golovach:016601,PhysRevLett.88.186802,PhysRevB.64.125316}.
Using $g^*=7$ and $J_{02}=1.1\,$meV we can overlay energies from a simple model of two-level repulsion to the current peaks in Fig.\,2 (green lines) with reasonable agreement and obtain a coupling matrix element \mbox{$\langle S(0,2)| H_{SO} | T_+(0,2)\rangle=\Delta_{SO}=0.2\,$meV}.

In conclusion we measured the HF and SO mixing energies of singlets and triplets in a weakly coupled InAs DQD in the regime of Pauli spin blockade. 
We were able to extract all relevant energy scales of the DQD and find a hierarchy \mbox{$J_{02}\gtrsim\Delta_{SO}>>t\gtrsim g^*\mu_\textrm{B}B_N$}. In contrast to DQDs in GaAs, SO interactions are efficient for small fields of a few mT. These energy scales suggest that InAs DQDs are suitable candidates for electric field induced spin manipulation \cite{golovach:165319,PhysRevLett.97.240501-flindt}
\begin{acknowledgments}
We thank T.~Ihn and J.~Taylor for stimulating discussions, M.~Borgstr\"om and E.~Gini for advice in nanowire growth. We acknowledge financial support from the ETH Zurich and I.S. thanks the European Commission for a Marie-Curie fellowship.
\end{acknowledgments}
\bibliography{MyBib.bib}

\begin{thebibliography}{29}
\expandafter\ifx\csname natexlab\endcsname\relax\def\natexlab#1{#1}\fi
\expandafter\ifx\csname bibnamefont\endcsname\relax
  \def\bibnamefont#1{#1}\fi
\expandafter\ifx\csname bibfnamefont\endcsname\relax
  \def\bibfnamefont#1{#1}\fi
\expandafter\ifx\csname citenamefont\endcsname\relax
  \def\citenamefont#1{#1}\fi
\expandafter\ifx\csname url\endcsname\relax
  \def\url#1{\texttt{#1}}\fi
\expandafter\ifx\csname urlprefix\endcsname\relax\def\urlprefix{URL }\fi
\providecommand{\bibinfo}[2]{#2}
\providecommand{\eprint}[2][]{\url{#2}}

\bibitem[{\citenamefont{Wolf et~al.}(2001)\citenamefont{Wolf, Awschalom,
  Buhrman, Daughton, von Moln\'ar, Roukes, Chtchelkanova, and
  Treger}}]{wolf2001}
\bibinfo{author}{\bibfnamefont{S.~A.} \bibnamefont{Wolf}},
  \bibinfo{author}{\bibfnamefont{D.~D.} \bibnamefont{Awschalom}},
  \bibinfo{author}{\bibfnamefont{R.~A.} \bibnamefont{Buhrman}},
  \bibinfo{author}{\bibfnamefont{J.~M.} \bibnamefont{Daughton}},
  \bibinfo{author}{\bibfnamefont{S.}~\bibnamefont{von Moln\'ar}},
  \bibinfo{author}{\bibfnamefont{M.~L.} \bibnamefont{Roukes}},
  \bibinfo{author}{\bibfnamefont{A.~Y.} \bibnamefont{Chtchelkanova}},
  \bibnamefont{and} \bibinfo{author}{\bibfnamefont{D.~M.}
  \bibnamefont{Treger}}, \bibinfo{journal}{Science}
  \textbf{\bibinfo{volume}{294}}, \bibinfo{pages}{1488} (\bibinfo{year}{2001}).

\bibitem[{\citenamefont{Datta and Das}(1990)}]{datta_das_SFET}
\bibinfo{author}{\bibfnamefont{S.}~\bibnamefont{Datta}} \bibnamefont{and}
  \bibinfo{author}{\bibfnamefont{B.}~\bibnamefont{Das}},
  \bibinfo{journal}{Appl. Phys. Lett.} \textbf{\bibinfo{volume}{56}},
  \bibinfo{pages}{665} (\bibinfo{year}{1990}).

\bibitem[{\citenamefont{Loss and DiVincenzo}(1998)}]{PhysRevA.57.120}
\bibinfo{author}{\bibfnamefont{D.}~\bibnamefont{Loss}} \bibnamefont{and}
  \bibinfo{author}{\bibfnamefont{D.~P.} \bibnamefont{DiVincenzo}},
  \bibinfo{journal}{Phys. Rev. A} \textbf{\bibinfo{volume}{57}},
  \bibinfo{pages}{120} (\bibinfo{year}{1998}).

\bibitem[{\citenamefont{Golovach et~al.}(2006)\citenamefont{Golovach, Borhani,
  and Loss}}]{golovach:165319}
\bibinfo{author}{\bibfnamefont{V.~N.} \bibnamefont{Golovach}},
  \bibinfo{author}{\bibfnamefont{M.}~\bibnamefont{Borhani}}, \bibnamefont{and}
  \bibinfo{author}{\bibfnamefont{D.}~\bibnamefont{Loss}},
  \bibinfo{journal}{Phys. Rev. B} \textbf{\bibinfo{volume}{74}},
  \bibinfo{eid}{165319} (\bibinfo{year}{2006});  
  \bibinfo{author}{\bibfnamefont{ D.~V.} \bibnamefont{Bulaev}} \bibnamefont{and}
  \bibinfo{author}{\bibfnamefont{D.}~\bibnamefont{Loss}},
  \bibinfo{journal}{Phys. Rev. Lett.} \textbf{\bibinfo{volume}{98}},
  \bibinfo{eid}{097202} (\bibinfo{year}{2007}).

\bibitem[{\citenamefont{Flindt et~al.}(2006)\citenamefont{Flindt, S\o{}rensen,
  and Flensberg}}]{PhysRevLett.97.240501-flindt}
\bibinfo{author}{\bibfnamefont{C.}~\bibnamefont{Flindt}},
  \bibinfo{author}{\bibfnamefont{A.~S.} \bibnamefont{S\o{}rensen}},
  \bibnamefont{and}
  \bibinfo{author}{\bibfnamefont{K.}~\bibnamefont{Flensberg}},
  \bibinfo{journal}{Phys. Rev. Lett.} \textbf{\bibinfo{volume}{97}},
  \bibinfo{pages}{240501} (\bibinfo{year}{2006}).

\bibitem[{\citenamefont{Kane}(1998)}]{KaneSiQC}
\bibinfo{author}{\bibfnamefont{B.~E.} \bibnamefont{Kane}},
  \bibinfo{journal}{Nature} \textbf{\bibinfo{volume}{393}},
  \bibinfo{pages}{133} (\bibinfo{year}{1998}).

\bibitem[{\citenamefont{Ono and Tarucha}(2004)}]{ono:256803}
\bibinfo{author}{\bibfnamefont{K.}~\bibnamefont{Ono}} \bibnamefont{and}
  \bibinfo{author}{\bibfnamefont{S.}~\bibnamefont{Tarucha}},
  \bibinfo{journal}{Phys. Rev. Lett.} \textbf{\bibinfo{volume}{92}},
  \bibinfo{eid}{256803} (\bibinfo{year}{2004}).

\bibitem[{\citenamefont{Erlingsson et~al.}(2001)\citenamefont{Erlingsson,
  Nazarov, and Fal\char39{}ko}}]{PhysRevB.64.195306}
\bibinfo{author}{\bibfnamefont{S.~I.} \bibnamefont{Erlingsson}},
  \bibinfo{author}{\bibfnamefont{Y.~V.} \bibnamefont{Nazarov}},
  \bibnamefont{and} \bibinfo{author}{\bibfnamefont{V.~I.}
  \bibnamefont{Falko}}, \bibinfo{journal}{Phys. Rev. B}
  \textbf{\bibinfo{volume}{64}}, \bibinfo{pages}{195306}
  (\bibinfo{year}{2001}).

\bibitem[{\citenamefont{Golovach et~al.}(2004)\citenamefont{Golovach,
  Khaetskii, and Loss}}]{golovach:016601}
\bibinfo{author}{\bibfnamefont{V.~N.} \bibnamefont{Golovach}},
  \bibinfo{author}{\bibfnamefont{A.}~\bibnamefont{Khaetskii}},
  \bibnamefont{and} \bibinfo{author}{\bibfnamefont{D.}~\bibnamefont{Loss}},
  \bibinfo{journal}{Phys. Rev. Lett.} \textbf{\bibinfo{volume}{93}},
  \bibinfo{eid}{016601} (\bibinfo{year}{2004}).

\bibitem[{\citenamefont{Sasaki et~al.}(2005)\citenamefont{Sasaki, Fujisawa,
  Hayashi, and Hirayama}}]{sasaki:056803}
\bibinfo{author}{\bibfnamefont{S.}~\bibnamefont{Sasaki}},
  \bibinfo{author}{\bibfnamefont{T.}~\bibnamefont{Fujisawa}},
  \bibinfo{author}{\bibfnamefont{T.}~\bibnamefont{Hayashi}}, \bibnamefont{and}
  \bibinfo{author}{\bibfnamefont{Y.}~\bibnamefont{Hirayama}},
  \bibinfo{journal}{Phys. Rev. Lett.} \textbf{\bibinfo{volume}{95}},
  \bibinfo{eid}{056803} (\bibinfo{year}{2005}).

\bibitem[{\citenamefont{Meunier et~al.}(2007)\citenamefont{Meunier, Vink, van
  Beveren, Tielrooij, Hanson, Koppens, Tranitz, Wegscheider, Kouwenhoven, and
  Vandersypen}}]{meunier:126601}
\bibinfo{author}{\bibfnamefont{T.}~\bibnamefont{Meunier}},
  \bibinfo{author}{\bibfnamefont{I.~T.} \bibnamefont{Vink}},
  \bibinfo{author}{\bibfnamefont{L.~H.} \bibnamefont{Willems van Beveren}},
  \bibinfo{author}{\bibfnamefont{K.-J.} \bibnamefont{Tielrooij}},
  \bibinfo{author}{\bibfnamefont{R.}~\bibnamefont{Hanson}},
  \bibinfo{author}{\bibfnamefont{F.~H.~L.} \bibnamefont{Koppens}},
  \bibinfo{author}{\bibfnamefont{H.~P.} \bibnamefont{Tranitz}},
  \bibinfo{author}{\bibfnamefont{W.}~\bibnamefont{Wegscheider}},
  \bibinfo{author}{\bibfnamefont{L.~P.} \bibnamefont{Kouwenhoven}},
  \bibnamefont{and} \bibinfo{author}{\bibfnamefont{L.~M.~K.}
  \bibnamefont{Vandersypen}}, \bibinfo{journal}{Phys. Rev. Lett.}
  \textbf{\bibinfo{volume}{98}}, \bibinfo{eid}{126601}
  (\bibinfo{year}{2007}).

\bibitem[{\citenamefont{Koppens et~al.}(2005)\citenamefont{Koppens, Folk,
  Elzerman, Hanson, van Beveren, Vink, Tranitz, Wegscheider, Kouwenhoven, and
  Vandersypen}}]{Koppens:2005qy}
\bibinfo{author}{\bibfnamefont{F.~H.~L.} \bibnamefont{Koppens}},
  \bibinfo{author}{\bibfnamefont{J.~A.} \bibnamefont{Folk}},
  \bibinfo{author}{\bibfnamefont{J.~M.} \bibnamefont{Elzerman}},
  \bibinfo{author}{\bibfnamefont{R.}~\bibnamefont{Hanson}},
  \bibinfo{author}{\bibfnamefont{L.~H.~W.} \bibnamefont{van Beveren}},
  \bibinfo{author}{\bibfnamefont{I.~T.} \bibnamefont{Vink}},
  \bibinfo{author}{\bibfnamefont{H.~P.} \bibnamefont{Tranitz}},
  \bibinfo{author}{\bibfnamefont{W.}~\bibnamefont{Wegscheider}},
  \bibinfo{author}{\bibfnamefont{L.~P.} \bibnamefont{Kouwenhoven}},
  \bibnamefont{and} \bibinfo{author}{\bibfnamefont{L.~M.~K.}
  \bibnamefont{Vandersypen}}, \bibinfo{journal}{Science}
  \textbf{\bibinfo{volume}{309}}, \bibinfo{pages}{1346} (\bibinfo{year}{2005}).

\bibitem[{\citenamefont{Johnson
  et~al.}(2005{\natexlab{a}})\citenamefont{Johnson, Petta, Taylor, Yacoby,
  Lukin, Marcus, Hanson, and Gossard}}]{johnsonPulse}
\bibinfo{author}{\bibfnamefont{A.~C.} \bibnamefont{Johnson}},
  \bibinfo{author}{\bibfnamefont{J.~R.} \bibnamefont{Petta}},
  \bibinfo{author}{\bibfnamefont{J.~M.} \bibnamefont{Taylor}},
  \bibinfo{author}{\bibfnamefont{A.}~\bibnamefont{Yacoby}},
  \bibinfo{author}{\bibfnamefont{M.~D.} \bibnamefont{Lukin}},
  \bibinfo{author}{\bibfnamefont{C.~M.} \bibnamefont{Marcus}},
  \bibinfo{author}{\bibfnamefont{M.~P.} \bibnamefont{Hanson}},
  \bibnamefont{and} \bibinfo{author}{\bibfnamefont{A.~C.}
  \bibnamefont{Gossard}}, \bibinfo{journal}{Nature}
  \textbf{\bibinfo{volume}{435}}, \bibinfo{pages}{925}
  (\bibinfo{year}{2005}{\natexlab{a}}).

\bibitem[{\citenamefont{Jouravlev and Nazarov}(2006)}]{Jouravlev:2006kx}
\bibinfo{author}{\bibfnamefont{O.~N.} \bibnamefont{Jouravlev}}
  \bibnamefont{and} \bibinfo{author}{\bibfnamefont{Y.~V.}
  \bibnamefont{Nazarov}}, \bibinfo{journal}{Phys. Rev. Lett.}
  \textbf{\bibinfo{volume}{96}}, \bibinfo{pages}{176804}
  (\bibinfo{year}{2006}).

\bibitem[{\citenamefont{Khaetskii et~al.}(2002)\citenamefont{Khaetskii, Loss,
  and Glazman}}]{PhysRevLett.88.186802}
\bibinfo{author}{\bibfnamefont{A.~V.} \bibnamefont{Khaetskii}},
  \bibinfo{author}{\bibfnamefont{D.}~\bibnamefont{Loss}}, \bibnamefont{and}
  \bibinfo{author}{\bibfnamefont{L.}~\bibnamefont{Glazman}},
  \bibinfo{journal}{Phys. Rev. Lett.} \textbf{\bibinfo{volume}{88}},
  \bibinfo{pages}{186802} (\bibinfo{year}{2002}).

\bibitem[{\citenamefont{Amasha et~al.}(2006)\citenamefont{Amasha, MacLean,
  Radu, Zumb\"uhl, Kastner, Hanson, and Gossard}}]{AmashaCondMat2006}
\bibinfo{author}{\bibfnamefont{S.}~\bibnamefont{Amasha}},
  \bibinfo{author}{\bibfnamefont{K.}~\bibnamefont{MacLean}},
  \bibinfo{author}{\bibfnamefont{I.}~\bibnamefont{Radu}},
  \bibinfo{author}{\bibfnamefont{D.~M.} \bibnamefont{Zumb\"uhl}},
  \bibinfo{author}{\bibfnamefont{M.~A.} \bibnamefont{Kastner}},
  \bibinfo{author}{\bibfnamefont{M.~P.} \bibnamefont{Hanson}},
  \bibnamefont{and} \bibinfo{author}{\bibfnamefont{A.~C.}
  \bibnamefont{Gossard}} (\bibinfo{year}{2006}),
  \bibinfo{note}{cond-mat/0607110}.

\bibitem[{\citenamefont{Khaetskii and Nazarov}(2001)}]{PhysRevB.64.125316}
\bibinfo{author}{\bibfnamefont{A.~V.} \bibnamefont{Khaetskii}}
  \bibnamefont{and} \bibinfo{author}{\bibfnamefont{Y.~V.}
  \bibnamefont{Nazarov}}, \bibinfo{journal}{Phys. Rev. B}
  \textbf{\bibinfo{volume}{64}}, \bibinfo{pages}{125316}
  (\bibinfo{year}{2001}).

\bibitem[{\citenamefont{Pfund et~al.}(2007)\citenamefont{Pfund, Shorubalko,
  Ensslin, and Leturcq}}]{pfund-2007-DNSP}
\bibinfo{author}{\bibfnamefont{A.}~\bibnamefont{Pfund}},
  \bibinfo{author}{\bibfnamefont{I.}~\bibnamefont{Shorubalko}},
  \bibinfo{author}{\bibfnamefont{K.}~\bibnamefont{Ensslin}}, \bibnamefont{and}
  \bibinfo{author}{\bibfnamefont{R.}~\bibnamefont{Leturcq}}
  (\bibinfo{year}{2007}), \bibinfo{note}{cond-mat/0701054}.

\bibitem[{\citenamefont{Fasth et~al.}(2007)\citenamefont{Fasth, Fuhrer,
  Samuelson, Golovach, and Loss}}]{fuhrer-SO-2007}
\bibinfo{author}{\bibfnamefont{C.}~\bibnamefont{Fasth}},
  \bibinfo{author}{\bibfnamefont{A.}~\bibnamefont{Fuhrer}},
  \bibinfo{author}{\bibfnamefont{L.}~\bibnamefont{Samuelson}},
  \bibinfo{author}{\bibfnamefont{V.~N.} \bibnamefont{Golovach}},
  \bibnamefont{and} \bibinfo{author}{\bibfnamefont{D.}~\bibnamefont{Loss}}
  (\bibinfo{year}{2007}), \bibinfo{note}{cond-mat/0701161}.

\bibitem[{\citenamefont{Pfund et~al.}(2006)\citenamefont{Pfund, Shorubalko,
  Leturcq, and Ensslin}}]{pfund:252106}
\bibinfo{author}{\bibfnamefont{A.}~\bibnamefont{Pfund}},
  \bibinfo{author}{\bibfnamefont{I.}~\bibnamefont{Shorubalko}},
  \bibinfo{author}{\bibfnamefont{R.}~\bibnamefont{Leturcq}}, \bibnamefont{and}
  \bibinfo{author}{\bibfnamefont{K.}~\bibnamefont{Ensslin}},
  \bibinfo{journal}{Appl. Phys. Lett.} \textbf{\bibinfo{volume}{89}},
  \bibinfo{eid}{252106} (\bibinfo{year}{2006}).

\bibitem[{\citenamefont{van~der Wiel et~al.}(2003)\citenamefont{van~der Wiel,
  Franceschi, Elzerman, Fujisawa, Tarucha, and Kouwenhoven}}]{vanderWiel01}
\bibinfo{author}{\bibfnamefont{W.~G.} \bibnamefont{van~der Wiel}},
  \bibinfo{author}{\bibfnamefont{S.~D.} \bibnamefont{Franceschi}},
  \bibinfo{author}{\bibfnamefont{J.~M.} \bibnamefont{Elzerman}},
  \bibinfo{author}{\bibfnamefont{T.}~\bibnamefont{Fujisawa}},
  \bibinfo{author}{\bibfnamefont{S.}~\bibnamefont{Tarucha}}, \bibnamefont{and}
  \bibinfo{author}{\bibfnamefont{L.~P.} \bibnamefont{Kouwenhoven}},
  \bibinfo{journal}{Rev. Mod. Phys.} \textbf{\bibinfo{volume}{75}},
  \bibinfo{eid}{1} (\bibinfo{year}{2003}).

\bibitem[{\citenamefont{Ono et~al.}(2002)\citenamefont{Ono, Austing, Tokura,
  and Tarucha}}]{Ono_rectification}
\bibinfo{author}{\bibfnamefont{K.}~\bibnamefont{Ono}},
  \bibinfo{author}{\bibfnamefont{D.~G.} \bibnamefont{Austing}},
  \bibinfo{author}{\bibfnamefont{Y.}~\bibnamefont{Tokura}}, \bibnamefont{and}
  \bibinfo{author}{\bibfnamefont{S.}~\bibnamefont{Tarucha}},
  \bibinfo{journal}{Science} \textbf{\bibinfo{volume}{297}},
  \bibinfo{pages}{1313} (\bibinfo{year}{2002}).

\bibitem[{\citenamefont{Johnson
  et~al.}(2005{\natexlab{b}})\citenamefont{Johnson, Petta, Marcus, Hanson, and
  Gossard}}]{johnson:165308}
\bibinfo{author}{\bibfnamefont{A.~C.} \bibnamefont{Johnson}},
  \bibinfo{author}{\bibfnamefont{J.~R.} \bibnamefont{Petta}},
  \bibinfo{author}{\bibfnamefont{C.~M.} \bibnamefont{Marcus}},
  \bibinfo{author}{\bibfnamefont{M.~P.} \bibnamefont{Hanson}},
  \bibnamefont{and} \bibinfo{author}{\bibfnamefont{A.~C.}
  \bibnamefont{Gossard}}, \bibinfo{journal}{Phys. Rev. B} \textbf{\bibinfo{volume}{72}}, \bibinfo{eid}{165308} (\bibinfo{year}{2005}{\natexlab{b}}).

\bibitem[{\citenamefont{Kouwenhoven et~al.}(2001)\citenamefont{Kouwenhoven,
  Austing, and Tarucha}}]{0034-4885-64-6-201}
\bibinfo{author}{\bibfnamefont{L.~P.} \bibnamefont{Kouwenhoven}},
  \bibinfo{author}{\bibfnamefont{D.~G.} \bibnamefont{Austing}},
  \bibnamefont{and} \bibinfo{author}{\bibfnamefont{S.}~\bibnamefont{Tarucha}},
  \bibinfo{journal}{Rep. Prog. Phys.}
  \textbf{\bibinfo{volume}{64}}, \bibinfo{pages}{701} (\bibinfo{year}{2001}).

\bibitem[{\citenamefont{Burkard et~al.}(1999)\citenamefont{Burkard, Loss, and
  DiVincenzo}}]{PhysRevB.59.2070}
\bibinfo{author}{\bibfnamefont{G.}~\bibnamefont{Burkard}},
  \bibinfo{author}{\bibfnamefont{D.}~\bibnamefont{Loss}}, \bibnamefont{and}
  \bibinfo{author}{\bibfnamefont{D.~P.} \bibnamefont{DiVincenzo}},
  \bibinfo{journal}{Phys. Rev. B} \textbf{\bibinfo{volume}{59}},
  \bibinfo{pages}{2070} (\bibinfo{year}{1999}).

\bibitem[{\citenamefont{Merkulov et~al.}(2002)\citenamefont{Merkulov, Efros,
  and Rosen}}]{PhysRevB.65.205309}
\bibinfo{author}{\bibfnamefont{I.~A.} \bibnamefont{Merkulov}},
  \bibinfo{author}{\bibfnamefont{A.~L.} \bibnamefont{Efros}}, \bibnamefont{and}
  \bibinfo{author}{\bibfnamefont{M.}~\bibnamefont{Rosen}},
  \bibinfo{journal}{Phys. Rev. B} \textbf{\bibinfo{volume}{65}},
  \bibinfo{pages}{205309} (\bibinfo{year}{2002}).

\bibitem[{\citenamefont{Goldman}(2005)}]{JRGoldmanThesis}
For the calculation in InAs we use the values given in 
\bibinfo{author}{\bibfnamefont{J.~R.} \bibnamefont{Goldman}},
  \bibinfo{type}{Phd thesis}, \bibinfo{school}{Stanford University}
  (\bibinfo{year}{2005}).

\bibitem[{\citenamefont{Dickmann and Hawrylak}(2003)}]{Dickmann:2003lr}
\bibinfo{author}{\bibfnamefont{S.}~\bibnamefont{Dickmann}} \bibnamefont{and}
  \bibinfo{author}{\bibfnamefont{P.}~\bibnamefont{Hawrylak}},
  \bibinfo{journal}{JETP Letters} \textbf{\bibinfo{volume}{V77}},
  \bibinfo{pages}{30} (\bibinfo{year}{2003}).

\bibitem[{\citenamefont{Tarucha et~al.}(1996)\citenamefont{Tarucha, Austing,
  Honda, van~der Hage, and Kouwenhoven}}]{PhysRevLett.77.3613}
\bibinfo{author}{\bibfnamefont{S.}~\bibnamefont{Tarucha}},
  \bibinfo{author}{\bibfnamefont{D.~G.} \bibnamefont{Austing}},
  \bibinfo{author}{\bibfnamefont{T.}~\bibnamefont{Honda}},
  \bibinfo{author}{\bibfnamefont{R.~J.} \bibnamefont{van~der Hage}},
  \bibnamefont{and} \bibinfo{author}{\bibfnamefont{L.~P.}
  \bibnamefont{Kouwenhoven}}, \bibinfo{journal}{Phys. Rev. Lett.}
  \textbf{\bibinfo{volume}{77}}, \bibinfo{pages}{3613} (\bibinfo{year}{1996}).

\end{thebibliography}
\end{document}